\documentclass[aps,prl,twocolumn,groupedaddress,showpacs]{revtex4}

\usepackage{amsmath}

\begin{document}



\title{Dynamical symmetry of Dirac hydrogen atom with spin symmetry and its connection with Ginocchio's oscillator}

\author{Fu-Lin Zhang}
\affiliation{Theoretical Physics Division, Chern Institute of
Mathematics, Nankai University, Tianjin 300071, People's Republic of
China}

\author{Bo Fu}
\affiliation{Theoretical Physics Division, Chern Institute of
Mathematics, Nankai University, Tianjin 300071, People's Republic of
China}

\author{Jing-Ling Chen}
\email[Email:]{chenjl@nankai.edu.cn}

\affiliation{Theoretical Physics Division, Chern Institute of
Mathematics, Nankai University, Tianjin 300071, People's Republic of
China}

\date{\today}

\begin{abstract}
The Dirac hydrogen atom with spin symmetry is shown has a $SO(4)$
symmetry. The generators are derived, and the corresponding Casimir
operator leads to the energy spectrum naturally. This type hydrogen
atom is connected to a four-dimensional Dirac system with equal
scalar and vector harmonic oscillator potential, by the
Kustaanheimo-Stiefel transformation with a constraint.
\end{abstract}


\maketitle

\textit{Introduction-}
 Hydrogen atom and harmonic oscillator are two
simple, solvable and real models. A common feature of them is that,
their orbits of motion are closed in classical mechanics
\cite{Bertand}. This indicates there are more constants of motion in
these systems than the orbit angular momentum. They have been shown
as the Rung-Lenz vector \cite{Lenz,Pauli} in the hydrogen atom and
the second order tensors \cite{Frakdin} in the harmonic oscillator.
These conserved quantities as well as angular momentum generate the
$SO(4)$ and $SU(3)$ Lie groups respectively. They are not
geometrical but the symmetries in the phase space, and are called
dynamical symmetries. These symmetries lead to an algebraic approach
to determine the energy levels. Generally, the $N-$dimensional (ND)
hydrogen atom has the $SO(N+1)$ and the isotropic oscillator has the
$SU(N)$ symmetry. The $SO(4)$ generators in the 3D hydrogen atom can
be written as two sets decoupled $SU(2)$ operators, whose Casimir
operators are equal \cite{Greiner}. On the other hand, one can
introduce a constraint condition on the 4D oscillator, and separate
it into two 2D oscillators with the same energy. These two systems
have the same algebraic structure, and can be connected by the
famous Kustaanheimo-Stiefel (KS) transformation \cite{KS}. The same
transformation has been applied to the path integral treatment of
the hydrogen atom \cite{Path} and the radial coherent state for the
Coulomb problem \cite{Coherent}.

In the relativistic quantum mechanics, the motion of spin-$1/2$
particle satisfies the Dirac equation. Neither the Dirac hydrogen
atom \cite{Greiner1} nor the Dirac oscillator \cite{DiracOsc} has a
dynamical symmetry. The main reason of the breaking of dynamical
symmetries is the spin-orbit coupling. There wasn't any
transformation reported to connect these two Dirac systems.

In recent years, the Dirac equation with scalar and vector
potentials of equal magnitude (SVPEM) are studied widely
\cite{ReSymm,Coulomb,Origin,Antinucleon,Hidden,Pseudospin,equivalent,equivalent2}.
When the potentials are spherical, the Dirac equation is said to
have the spin or pseudospin symmetry corresponding to the same or
opposite sign. Then, the total angular momentum can be divided into
conserved orbital and spin parts (see Eq. (\ref{L&S}) for the spin
symmetry case), which form the $SU(2)$ algebra separately. These
symmetries, which have been observed in the hadron and nuclear
spectroscopies for a long time \cite{observe,observe2}, are derived
from the investigation of the dynamics between a quark and an
antiquark \cite{quark,LS,Crater}. The very lately studies
\cite{equivalent,equivalent2} have revealed that, the motion of a
spin-1/2 particle with SVPEM satisfies the same differential
equation and has the same energy spectrum as a scalar particle. For
the spin or pseudospin symmetry systems, Alberto \textit{et. al.}
\cite{equivalent} have indicated that the spin-orbit and Darwin
terms of either the upper component or the lower component of the
Dirac spinor vanish, which made it equivalent, as far as energy is
concerned, to a spin-0 state. These results suggest that, one can
image the spin-1/2 particle with SVPEM as a relativistic scalar
particle with an additional spin but without the spin-orbit
coupling.  From this point of view, we speculate, the dynamical
symmetries in non-relativistic hydrogen atom and harmonic oscilltor
would be holden in the Dirac systems with spin or pseudospin
symmetry.

Actually, Ginocchio \cite{U31,U32} has found the $U(3)$ and
pseudo-$U(3)$ symmetry in the Dirac equation with spin or pseudospin
symmetry when the potential takes the harmonic oscillator form. The
aim of this work is to answer the following questions: Does the
Dirac equation with spin or pseudospin symmetry has a $SO(4)$
dynamical symmetry, when the potential takes the Coulomb form? And
what is the relation of it and the Ginocchio's oscillator? For the
sake of brevity, we only given the details of the spin symmetry
case. The results of the pseudospin symmetry case will be easily
obtained by making some corrections.

\textit{SO(4) symmetry-}
 The Dirac Hamiltonian with spin symmetry,
in the relativistic units, $\hbar=c=1$, is given by
\begin{eqnarray}\label{H}
H=\vec{\alpha}\cdot\vec{p}+\beta M +(1+\beta)\frac{V(r)}{2},
\end{eqnarray}
where $\vec{\alpha}$ and $\beta$ are the Dirac matrices, and $M$ is
the mass. It commutes with the deformed orbital and spin angular
momentum \cite{LS}
\begin{eqnarray}\label{L&S}
\ \vec{L}=\begin{bmatrix}
 \vec{l}&0\\
 0& U_p\vec{l}U_p^{\dag}
 \end{bmatrix},
\ \ \vec{S}=\begin{bmatrix}
 \vec{s}&0\\
 0& U_p\vec{s}U_p^{\dag}
 \end{bmatrix},
 \end{eqnarray}
where $\vec{l}=\vec{r} \times \vec{p}$,
$\vec{s}=\frac{\vec{\sigma}}{2}$ are the usual spin generators,
$\vec{\sigma}$ are the Pauli matrices, and
$U_p=U_p^{\dag}=\frac{\vec{\sigma}\cdot\vec{p}}{p}$ is the helicity
unitary operator \cite{Hel}. The components of them form the $SU(2)$
Lie algebra separately, and the sum of them equals to the total
angular momentum on account of
$U_p\vec{l}U_p^{\dag}+U_p\vec{s}U_p^{\dag}=\vec{l}+\vec{s}$.
Ginocchio has shown $\vec{L}$ in Eq. (\ref{L&S}) to be three of the
generators of the $U(3)$ symmetry group. The other conserved
quantities are supposed to take the form as
\begin{eqnarray}\label{Q}
\ Q=\begin{bmatrix}
 Q_{11} & Q_{12} \vec{\sigma}\cdot\vec{p} \\
  \vec{\sigma}\cdot\vec{p} Q_{21} \ &
  \vec{\sigma}\cdot\vec{p} Q_{22} \vec{\sigma}\cdot\vec{p}
 \end{bmatrix}. \nonumber
\end{eqnarray}
If it commutes with the spin symmetry Hamiltonian (\ref{H}), the
matrix elements should satisfy the equations
\begin{eqnarray}\label{Qeqn}
Q_{12} &=& Q_{21},\nonumber\\
\lbrack Q_{11},V(r)\rbrack +\lbrack Q_{12},p^2 \rbrack   &=& 0,\\
\lbrack Q_{12},V(r)\rbrack +\lbrack Q_{22},p^2 \rbrack   &=& 0,\nonumber\\
Q_{11} &=& Q_{12}(2M+V(r))+Q_{22}p^2.\nonumber
\end{eqnarray}

In non-relativistic hydrogen atom, the constants of motion are the
orbital angular momentum $\vec{l}$ and the Rung-Lenz vector
\cite{Lenz,Pauli,Greiner},
$\vec{R}=\frac{\vec{f}}{2Mk}-\frac{\vec{r}}{r}$,
where $\vec{f}=\vec{p} \times \vec{l}- \vec{l} \times \vec{p}$, and
$k$ is the parameter in the Coulomb potential $V(r)^h=-\frac{k}{r}$.
One can obtain the following relations easily
\begin{eqnarray}\label{CR}
\lbrack \vec{f},p^2 \rbrack = 0,\ \ \
\lbrack -\frac{\vec{r}}{r},V^h(r) \rbrack = 0,\nonumber\\
\frac{1}{2Mk}\lbrack \vec{f},V^h(r) \rbrack + \frac{1}{2M} \lbrack
-\frac{\vec{r}}{r},p^2 \rbrack = 0.\nonumber
\end{eqnarray}
Insert them into Eq. (\ref{Qeqn}), we can find the solutions
 of the Coulomb potential $V(r)=V^h(r)=-\frac{k}{r}$
\begin{eqnarray}\label{QH}
\ \vec{Q}=\begin{bmatrix}
 \ 2M \vec{R} +\frac{k \vec{r}}{r^2}&(-\frac{\vec{r}}{r})\vec{\sigma}\cdot\vec{p}\\
 \vec{\sigma}\cdot\vec{p}(-\frac{\vec{r}}{r})& \vec{\sigma}\cdot\vec{p}(\frac{1}{k}\frac{\vec{f}}{p^2})\vec{\sigma}\cdot\vec{p}
 \end{bmatrix}. \nonumber
\end{eqnarray}
The commutation relations of conserved $\vec{L}$ and $\vec{Q}$ are
\begin{eqnarray}\label{LR}
\lbrack L_i,L_j \rbrack &=& i \epsilon_{ijk} L_k,\nonumber\\
\lbrack L_i,Q_j \rbrack &=& i \epsilon_{ijk} Q_k,  \\
\lbrack Q_i,Q_j \rbrack &=& i \frac{-4}{k^2} (H^2-M^2)
\epsilon_{ijk} L_k,\nonumber
\end{eqnarray}
 where $i,j,k=1,2,3$. And,
\begin{eqnarray}\label{QL}
&&\vec{Q} \cdot \vec{L}  = 0, \nonumber\\
&&{\vec{Q}}^2 = \frac{4}{k^2}(H^2-M^2)(L^2+1)+(H+M)^2.
\end{eqnarray}
 For a fixed energy level, $H$ is a constant. We can define the
normalized generators $\vec{A}= [\frac{-4}{k^2}
(H^2-M^2)]^{-\frac{1}{2}} \vec{Q}$. Evidently, the two later
relations in Eq. (\ref{LR}) are transformed into $\lbrack L_i,A_j
\rbrack = i \epsilon_{ijk} A_k$ and $\lbrack A_i,A_j \rbrack = i
\epsilon_{ijk} L_k$. These results show that the Dirac hydrogen atom
with spin symmetry has a $SO(4)$ symmetry.

By the standard process of the non-relativistic hydrogen atom
\cite{Greiner}, the relations of the generators lead to the energy
spectrum. Set $\vec{I}=(\vec{L} + \vec{A})/2$ and $\vec{K}=(\vec{L}
- \vec{A})/2$, which satisfy the commutation relations, $\lbrack
I_i,I_j \rbrack= i \epsilon_{ijk} I_k,\ \lbrack K_i,K_j \rbrack= i
\epsilon_{ijk} K_k,\ \lbrack I_i,K_j \rbrack= 0$.
$\vec{I}$ and $\vec{K}$ constitute two commutative $SU(2)$ algebra.
Considering Eq. (\ref{QL}), it is immediate to obtain the Casimir
operators
\begin{eqnarray}\label{Csu2}
{\vec{I}}^2={\vec{K}}^2=\frac{1}{4} \bigr[ -\frac{k^2}{4}
\frac{H+M}{H-M} -1 \bigr] =j(j+1), \nonumber
\end{eqnarray}
where $j=0,1/2,1,3/2...$ Hence, the eigenvalues of the Hamiltonian
are given by
\begin{eqnarray}\label{Eh}
E^{\pm}=\frac{\pm 4n^2-k^2}{4n^2+k^2}M,\ \ \ n=2j+1=1,2,3...
\end{eqnarray}
which come up to \cite{ReSymm,Coulomb}.

When $M \rightarrow \infty$, $H \rightarrow M$, the non-relativistic
limit of the energy levels is $E^{+} \rightarrow
M-\frac{k^2}{2n^2}M$,
the second term of which agrees with the non-relativistic results
\cite{Greiner}. We can also get the non-relativistic limits of the
conserved quantities
\begin{eqnarray}\label{LmtQHh}
H-M  \rightarrow
\begin{bmatrix}
 \frac{p^2}{2M}-\frac{k}{r}&0\\
 0&\frac{p^2}{2M}.
 \end{bmatrix}, \ \
\frac{\vec{Q}}{2M}  \rightarrow  \begin{bmatrix}
 \ \vec{R} &0\\
 0&\frac{1}{2Mk} \vec{f}
 \end{bmatrix}. \nonumber
\end{eqnarray}
The upper-left elements of the above matrices are nothing but the
non-relativistic hydrogen atom Hamiltonian and the Lung-Lenz vector,
and the lower-right ones are their limits when $k \rightarrow 0$.

\textit{KS transformation-}
 The treatment of dynamical symmetry in
the above section can be generalized to ND Dirac systems with SVPEM.
This will be discussed in a forthcoming paper. Here we give the
spectrum of the 4D Dirac system with equal scalar and vector
harmonic oscillator potentials (4D Ginocchio's oscillator), and show
its relation with the Dirac hydrogen atom with spin symmetry.
Denoting the 4D spatial coordinates and momentums as $u_{\mu}$ and $
P_{\mu} (\mu=1,2,3,4)$,
 the Dirac Hamiltonian with SVPEM is given by
\begin{eqnarray}\label{H4}
\mathcal{H}=\alpha^{\mu} P_{\mu}+ \beta m
+(1+\beta)\frac{\mathcal{V}(u)}{2}. \nonumber
\end{eqnarray}
Here, $\alpha^{1,2,3}=\vec{\alpha}$ and $\beta$ are the same as the
3D case, and $\alpha^{4}=- \sigma_2 \otimes I$. Then, the Dirac
equation can be written as
\begin{eqnarray}\label{DE4}
\begin{bmatrix}
 \ m+\mathcal{V}(u)-\varepsilon&B\\
 B^{\dag}&-m-\varepsilon
\end{bmatrix}  \begin{bmatrix}
 \phi_1\\
 \phi_2
\end{bmatrix}=0,
\end{eqnarray}
where $B=\sigma^{\mu} P_{\mu}$ with $\sigma^{1,2,3}=\vec{\sigma}$
and $\sigma^{4}=i$. It equals to the second order differential
equations
\begin{eqnarray}
&&[P^2+(\varepsilon+m)\mathcal{V}(u)-(\varepsilon^2-m^2)]\phi_1=0, \nonumber\\
&&(\varepsilon+m) \phi_2 + B^{\dag} \phi_1 = 0. \nonumber
\end{eqnarray}
Considering $\varepsilon$ is a constant for a given energy level,
the first equation takes the same form as the Schr\"{o}dinger
equation with the mass $\tilde{m}=(\varepsilon+m)/2$ and eigenvalue
$\tilde{\varepsilon}=\varepsilon-m$. When
$\mathcal{V}(u)=\frac{1}{2}m \omega^2 u^2 =\frac{1}{2}\tilde{m}
\Omega^2 u^2$, the energy spectrum of the 4D Schr\"{o}dinger
equation is $\tilde{\varepsilon}=(N+2)\Omega$, where
$N=n_1+n_2+n_3+n_4$ is the total number operator \cite{Coherent} of
the 4D non-relativistic harmonic oscillator. Hence, the eigenvalues
in Eq. (\ref{DE4}) satisfy the equation
\begin{eqnarray}\label{Eeqn}
(\varepsilon+m)^2(\varepsilon-m)^2-{2m
\omega^2}(\varepsilon+m)(N+2)^2=0.
\end{eqnarray}

The so-called KS transformation, between the spatial coordinates
$\vec{r}=(x_1,x_2,x_3)$ and $u_{\mu}=(u_1,u_2,u_3,u_4)$, is given by
\cite{KS}
\begin{eqnarray} \label{KS}
x_1&=&2(u_1 u_3 -u_2 u_4),\nonumber\\
x_2&=&2(u_1 u_4 +u_2 u_3), \nonumber \\
x_3&=&u_1^2 + u_2^2 -u_3^2 - u_4^2,\nonumber
\end{eqnarray}
which connects the 3D hydrogen atom and 4D harmonic oscillator with
the constraint condition
\begin{eqnarray} \label{Constr}
u_4 P_1-u_1P_4+u_2 P_3 -u_3 P_2=0.
\end{eqnarray}
 Under this transformation and constraint, it is easy to obtain $B=2 \Gamma
\vec{\sigma} \cdot \vec{p}$, where $\Gamma=u_1 + i u_2 \sigma_3 - i
u_3 \sigma_2 + i u_4 \sigma_1$, $\Gamma \Gamma^{\dag}=u^2=r$. Then,
the Dirac equation (\ref{DE4}) can be written as
\begin{eqnarray} \label{DE4to3}
\begin{bmatrix}
 \ 2\Gamma&0\\
 0&1
\end{bmatrix}
\begin{bmatrix}
 \ \frac{m-\varepsilon}{4r} + \frac{1}{8}m \omega^2& \vec{\sigma} \cdot \vec{p}\\
 \vec{\sigma} \cdot \vec{p} &-m-\varepsilon
\end{bmatrix}
\begin{bmatrix}
 \ 2\Gamma^{\dag}&0\\
 0&1
\end{bmatrix}  \begin{bmatrix}
 \phi_1\\
 \phi_2
\end{bmatrix}=0.\ \
\end{eqnarray}
Setting
\begin{eqnarray}\label{relation}
M+E=m+\varepsilon,\ \ M-E=\frac{1}{8}m \omega^2 ,\ \
k=\frac{1}{4}(\varepsilon-m), \ \
\end{eqnarray}
Eq. (\ref{DE4to3}) is equivalent to nothing but the 3D Dirac
equation of the hydrogen atom with spin symmetry
\begin{eqnarray}
\begin{bmatrix}
 \ M-\frac{k}{r}-E & \vec{\sigma} \cdot \vec{p}\\
 \vec{\sigma} \cdot \vec{p} &-M-E
\end{bmatrix}
\begin{bmatrix}
 \psi_1\\
 \psi_2
\end{bmatrix}=0,
\end{eqnarray}
where $\psi_1=2 \Gamma^{\dag} \phi_1$ and $\psi_2= \phi_2$. The
constraint in Eq. (\ref{Constr}) requires the number operators in 4D
Schr\"odinger harmonic oscillator satisfy \cite{Coherent} $
n_1+n_2=n_3+n_4$. Therefore, the quantum number in Eq. (\ref{Eeqn})
is confined to be even, $N+2=2n=2,4,6...$ Substituting the relations
from Eq. (\ref{relation}) into Eq. (\ref{Eeqn}), one can obtain the
energy spectrum in Eq. (\ref{Eh}).

\textit{Conclusion-}
 In summary, we have shown that the 3D Dirac
hydrogen atom with spin symmetry has the  $SO(4)$ symmetry. The
nature of this symmetry is not geometrical but dynamical. The
relation of its Hamiltonian and the Casimir operators of the
symmetry group leads to an algebraic solution of the relativistic
energy spectrum. The non-relativistic limits of the conserved
quantities coincide with their non-relativistic counterpart
accurately. As a result of the symmetries of the Coulomb and
harmonic oscillator problem, we prove the Dirac hydrogen atom with
spin symmetry can be connected to a 4D Ginocchio's oscillator by the
KS transformation with a constraint condition. Dynamical symmetry
and the KS transformation are two important concepts in
non-relativistic quantum mechanics. The results in this paper and
\cite{U32} show they also exist in Dirac systems with SVPEM. Since
the spin symmetry and pseudospin symmetry exist frequently in
antinucleon and nucleon spectrum \cite{ReSymm}. We can foretell
there are more properties of the Schr\"odinger equations presence in
these systems.

\begin{acknowledgments}
This work is supported in part by NSF of China (Grants No. 10575053
and No. 10605013) and Program for New Century Excellent Talents in
University. The Project-sponsored by SRF for ROCS, SEM.
\end{acknowledgments}

\bibliography{DiarcHA}

\end{document}